\documentstyle[12pt]{article}
\input psfig
\catcode`\@=11
\long\def\@makefntext#1{
\protect\noindent \hbox to 3.2pt {\hskip-.9pt
$^{{\ninerm\@thefnmark}}$\hfil}#1\hfill}                

\def\@makefnmark{\hbox to 0pt{$^{\@thefnmark}$\hss}}  

\def\ps@myheadings{\let\@mkboth\@gobbletwo
\def\@oddhead{\hbox{}
\rightmark\hfil\ninerm\thepage}
\def\@oddfoot{}\def\@evenhead{\ninerm\thepage\hfil
\leftmark\hbox{}}\def\@evenfoot{}
\def\sectionmark##1{}\def\subsectionmark##1{}}

\renewcommand{\thefootnote}{\fnsymbol{footnote}}

\newcounter{sectionc}\newcounter{subsectionc}\newcounter{subsubsectionc}
\renewcommand{\section}[1] {\vspace*{0.6cm}\addtocounter{sectionc}{1}
\setcounter{subsectionc}{0}\setcounter{subsubsectionc}{0}\noindent
	{\normalsize\bf\thesectionc. #1}\par\vspace*{0.4cm}}
\renewcommand{\subsection}[1] {\vspace*{0.6cm}\addtocounter{subsectionc}{1}
	\setcounter{subsubsectionc}{0}\noindent
	{\normalsize\it\thesectionc.\thesubsectionc. #1}\par\vspace*{0.4cm}}
\renewcommand{\subsubsection}[1]
{\vspace*{0.6cm}\addtocounter{subsubsectionc}{1}
	\noindent {\normalsize\rm\thesectionc.\thesubsectionc.\thesubsubsectionc.
	#1}\par\vspace*{0.4cm}}

\newcounter{appendixc}
\newcounter{subappendixc}[appendixc]
\newcounter{subsubappendixc}[subappendixc]

\renewcommand{\appendix}[1] {\vspace*{0.6cm}
	\refstepcounter{appendixc}
	\setcounter{figure}{0}
	\setcounter{table}{0}
	\setcounter{equation}{0}
	\renewcommand{\thefigure}{\Alph{appendixc}.\arabic{figure}}
	\renewcommand{\thetable}{\Alph{appendixc}.\arabic{table}}
	\renewcommand{\theappendixc}{\Alph{appendixc}}
	\renewcommand{\theequation}{\Alph{appendixc}.\arabic{equation}}
	\noindent{\bf Appendix \theappendixc #1}\par\vspace*{0.4cm}}

\def\abstracts#1{{

\centering{\begin{minipage}{12.2truecm}\footnotesize\baselineskip=12pt\noindent
	\centerline{\footnotesize ABSTRACT}\vspace*{0.3cm}
	\parindent=0pt #1
	\end{minipage}}\par}}

\renewenvironment{thebibliography}[1]
	{\begin{list}{\arabic{enumi}.}
	{\usecounter{enumi}\setlength{\parsep}{0pt}
\setlength{\leftmargin 1.25cm}{\rightmargin 0pt}
	 \setlength{\itemsep}{0pt} \settowidth
	{\labelwidth}{#1.}\sloppy}}{\end{list}}

\newcounter{itemlistc}
\newcounter{romanlistc}
\newcounter{alphlistc}
\newcounter{arabiclistc}

\newcommand{\fcaption}[1]{
	\refstepcounter{figure}
	\setbox\@tempboxa = \hbox{\footnotesize Fig.~\thefigure. #1}
	\ifdim \wd\@tempboxa > 6in
	   {\begin{center}
	\parbox{6in}{\footnotesize\baselineskip=12pt Fig.~\thefigure. #1}
	    \end{center}}
	\else
	     {\begin{center}
	     {\footnotesize Fig.~\thefigure. #1}
	      \end{center}}
	\fi}

\newcommand{\tcaption}[1]{
	\refstepcounter{table}
	\setbox\@tempboxa = \hbox{\footnotesize Table~\thetable. #1}
	\ifdim \wd\@tempboxa > 6in
	   {\begin{center}
	\parbox{6in}{\footnotesize\baselineskip=12pt Table~\thetable. #1}
	    \end{center}}
	\else
	     {\begin{center}
	     {\footnotesize Table~\thetable. #1}
	      \end{center}}
	\fi}

\def\@citex[#1]#2{\if@filesw\immediate\write\@auxout
	{\string\citation{#2}}\fi
\def\@citea{}\@cite{\@for\@citeb:=#2\do
	{\@citea\def\@citea{,}\@ifundefined
	{b@\@citeb}{{\bf ?}\@warning
	{Citation `\@citeb' on page \thepage \space undefined}}
	{\csname b@\@citeb\endcsname}}}{#1}}

\newif\if@cghi
\def\cite{\@cghitrue\@ifnextchar [{\@tempswatrue
	\@citex}{\@tempswafalse\@citex[]}}
\def\citelow{\@cghifalse\@ifnextchar [{\@tempswatrue
	\@citex}{\@tempswafalse\@citex[]}}
\def\@cite#1#2{{$\null^{#1}$\if@tempswa\typeout
	{IJCGA warning: optional citation argument
	ignored: `#2'} \fi}}

 1
 1
 1

\font\ninerm=cmr9

\newcommand{\eq}[1]{(\ref{#1})}
\newcommand{\diff}{\partial}
\newcommand{\beq}{\begin{equation}}
\newcommand{\eeq}{\end{equation}}
\newcommand{\beqn}{\begin{eqnarray}}
\newcommand{\eeqn}{\end{eqnarray}}

\def\cf{{\it cf.}}

\newcommand{\cZ}{{\cal Z}}

\def\dd{{\rm d}}  
\def\NP{ Nucl.~Phys.}
\def\PL{ Phys.~Lett.}

\def\PR{ Phys.~Rev.}
\date{}
\begin{document}

\begin{flushright}
ITEP-95-25\\
hep-lat/9504013
\end{flushright}
\vspace{1.7cm}

\centerline{\normalsize\bf TOPOLOGICAL OBJECTS AND CONFINEMENT}
\baselineskip=22pt
\centerline{\normalsize\bf
ON THE LATTICE\footnote{Talk given by M.I.Polikarpov at International
RCNP Workshop on Color Confinement and Hadrons (Confinement 95),
March 22--24, 1995, RCNP, Osaka, Japan}
}
\vspace*{0.6cm}
\centerline{\footnotesize E.T.~AKHMEDOV,}
\baselineskip=13pt
\centerline{\footnotesize\it ITEP, B.Cheremushkinskaya 25, Moscow,
117259, Russia}
\vspace*{0.3cm}
\centerline{\footnotesize M.N.~CHERNODUB}
\baselineskip=13pt
\centerline{\footnotesize\it ITEP, B.Cheremushkinskaya 25, Moscow,
117259, Russia}
\vspace*{0.3cm}
\centerline{\footnotesize and}
\vspace*{0.3cm}
\centerline{\footnotesize M.I.~POLIKARPOV}
\baselineskip=13pt
\centerline{\footnotesize\it ITEP, B.Cheremushkinskaya 25, Moscow,
117259, Russia}
\baselineskip=12pt
\centerline{\footnotesize E-mail: polykarp@vxdesy.desy.de}
\vspace*{0.9cm}
\abstracts{First we discuss various topological objects (monopoles,
``minopoles'' and ``hybrids'') which may be important for the confinement
mechanism in various abelian projections. The second topic is the
string between quark and antiquark. The standard quantum string with the
Nambu-Goto action exists only in D=26. If we start from the field theory, in
which the string excitations exist, and change the variables in the path
integral to the string variables, then the Jacobian appears. This Jacobian
generates the correction to the Nambu-Goto action. For this effective
action the conformal anomaly cancels in D=4. Thus we get the quantum string
theory in D=4.}

\normalsize\baselineskip=15pt
\setcounter{footnote}{0}
\renewcommand{\thefootnote}{\alph{footnote}}
\vfill
\pagebreak

\section{Introduction}
Many numerical experiments confirm the monopole confinement
mechanism~\cite{Man76,tHo76} in the $U(1)$ theory obtained by the abelian
projection~\cite{tHo81} from the $SU(2)$ lattice gluodynamics. The well
known examples are:

\begin{itemize}
\item
The string tension $\sigma_{U(1)}$ calculated from the $U(1)$ Wilson loops
(loops constructed only from the abelian gauge fields) coincides with the
full $SU(2)$ string tension~\cite{SuYo90}.

\item The density of the monopoles seems to scale~\cite{Bor91}.

\item The monopole currents satisfy the
London equation for a superconductor~\cite{SiBrHa93}.

\item
The $SU(2)$ string tension can be obtained, with good accuracy,
from the contribution of the abelian monopole
currents~\cite{ShSu94,StNeWe94}.

\end{itemize}

All these remarkable facts, however, have been obtained only
for the so called maximal abelian (MaA) projection
\cite{KrScWi87,KrLaScWi87}. Other abelian projections (such as the
diagonalization of the plaquette matrix $U_{x,12}$) do not give evidence
that the vacuum behaves as the dual superconductor\footnote{In the talk of
A.~Di Giacomo at this workshop (see also\cite{DeGiPa94}) it has been
claimed that the value of the monopole condensate is the order parameter for
the phase transition. This result is obtained for the abelian projection in
which the Polyakov line is diagonalized; therefore there is a
similarity between this gauge and the MaA projection.}. Below we give three
examples.

	First, it turns out~\cite{IvPoPo90} that the fractal dimensionality
of the monopole currents extracted from the lattice vacuum by means of the
maximal abelian projection is strongly correlated with the string tension.
If monopoles are extracted by means of other projections, this correlation is
absent (\cf\ Fig.2 and Fig.4 of ref.\cite{IvPoPo90}). Another example is the
temperature dependence of the monopole condensate measured on the basis of
the percolation properties of the clusters of monopole currents
\cite{IvPoPo93}. For the maximal abelian projection the condensate is
nonzero below the critical temperature $T_c$ and vanishes above it. For the
projection which corresponds to the diagonalization of $U_{x,12}$, the
condensate is nonzero at $T>T_c$, and it is not the order parameter for the
phase transition. The last result has been obtained by the authors of
\cite{IvPoPo93}, but has not been published. The space--time asymmetry of the
monopole currents behaves as the order parameter for the deconfinement phase
transition for the MaA projection, and is zero both below and above the
critical temperature for the so called minimal Abelian projection
\cite{ChPoVe95}. In Section 2 we discuss the dependence of the
confinement mechanism on the type of the abelian projection.
In Section 3 we show what kind of quantum strings may exist between the quark
and the antiquark.

\section{Abelian Projection of the $SU(2)$ Gauge Theory}

After the Abelian projection of the $SU(2)$ gluodynamics,
the diagonal elements of the gauge field become the $U(1)$ gauge field, and
the nondiagonal elements become charged matter fields; this is clear from
the $U(1)$ gauge transformations: $A^{ii}_\mu \rightarrow A^{ii}_\mu\, +
\,\diff_\mu\alpha$, $A^{\pm}_\mu \rightarrow
A^{\pm}_\mu \, e^{\pm 2 i \alpha}$. On the lattice, for the standard
parametrization of the link matrix, $U^{11}_l= \cos \phi e^{i\theta},$
$U^{12}_l = \sin \phi e^{i\chi}$, the situation is similar. After the Abelian
projection, $\theta$ becomes the compact abelian gauge field; and $\chi$
becomes the compact matter field. The $U(1)$ gauge transformations are:

\beq
\theta \rightarrow \theta + \alpha_1 -\alpha_2, \label{u1th}
\eeq

\beq
\chi \rightarrow \chi + \alpha_1 + \alpha_2. \label{u1chi}
\eeq

\subsection{Maximal and Minimal Abelian projections}
The widely used MaA projection~\cite{KrScWi87,KrLaScWi87} corresponds to the
gauge transformation that makes the link matrices diagonal ``as much as
possible''. For the $SU(2)$ lattice gauge theory, the matrices of the gauge
transformation $\Omega_x$ are defined by the following maximization
condition:

\beqn
\max_{\{\Omega_x\}}R(U')\;,\ R(U') = \sum_{x,\mu} \, Tr(U'_{x\mu}\sigma_3
U'^{+}_{x\mu}\sigma_3), \label{MaA}\\
U'_{x\mu} = \Omega^+_x U_{x\mu} \Omega_{x+\hat{\mu}}\;.\nonumber
\eeqn

In order to show that in discussing of the confinement mechanism one has to
take into account the type of the Abelian projection, we consider the
Minimal Abelian (MiA) projection~\cite{ChPoVe95} defined as:

\beq
     \min_{\{\Omega_x\}} R(U')\;, \label{MiA}
\eeq

The action of the $SU(2)$ gluodynamics can be represented in the following
form:

\beqn
S=\beta\ \, \mbox{Tr} \, U_P
= \beta_1(\phi) \cos\theta_P + S^{int}_\phi(\theta,\chi)
+ \beta_2(\phi) \cos\chi_{\tilde{P}} \label{S}
\eeqn

For the MaA projection, $\beta_1$ is large, the first term in the sum
dominates, and if we neglect fluctuations of the angle
$\phi$, as well as the Faddeev-Popov determinant, the $SU(2)$ action in the
MaA projection is well approximated by the $U(1)$ action: $S_P\approx
\bar{\beta}\cos \theta_P$, $\bar{\beta}=\beta <\cos \phi>^4$. The fields in
the MaA projection can be transformed into those
in the MiA projection by the gauge transformation, and the roles of
the fields $\theta$ and $\chi$ are interchanged~\cite{ChPoVe95}. For the
MiA projection $\beta_2$ is large, and the gluodynamics is
approximately reduced
to the theory of the vector matter field $\chi$:  $S \approx \beta_2(\phi)
\cos\chi_{\tilde{P}}$.

\subsection{Monopoles and Minopoles}
The monopoles extracted from the
field $\theta$ in the MaA projection turn, in the MiA projection,
into certain
topological defects constructed from the ``matter'' fields $\chi$. We call
these topological defects ``minopoles''.
Minopoles can be extracted from a given configuration of gauge fields
similarly to monopoles: from the angles $\chi$ we construct the $U(1)$
invariant plaquette variables $\chi_{\tilde{P}} =
\chi_1-\chi_2+\chi_3-\chi_4 \bmod 2\pi$. From these plaquette
variables we construct the variables attached to the elementary cubes ${}^*j
= \frac{1}{2\pi}\tilde{\dd} \chi_{\tilde{P}}$; for ${}^*j \neq 0$ the link
dual to the cube carries the minopole current. We use the notation
$\tilde{\dd}$ (instead of $\dd$), since the gauge transformations of $\chi$
given by \eq{u1chi} differ from the gauge transformations of $\theta$ given
by \eq{u1th}, and the construction of the plaquette variable from the link
variables and the construction
of the cube variable from the plaquette variables differ
in an obvious way from the standard construction.
In Fig.1(a)
we illustrate the standard construction of the monopoles from the field
$\theta$, and in Fig.1(b) we show the construction of the minopoles from the
field $\chi$. The variables $\theta$ are characterized by the direction,
shown by arrows; the variables $\chi$ are characterized by their
sign, and the variables $\chi$ which transforms as in eq.
\eq{u1chi} are shown by solid lines; the variables $\chi$ with the
opposite sign are shown by the dashed lines.

\begin{figure}
\psfig{file=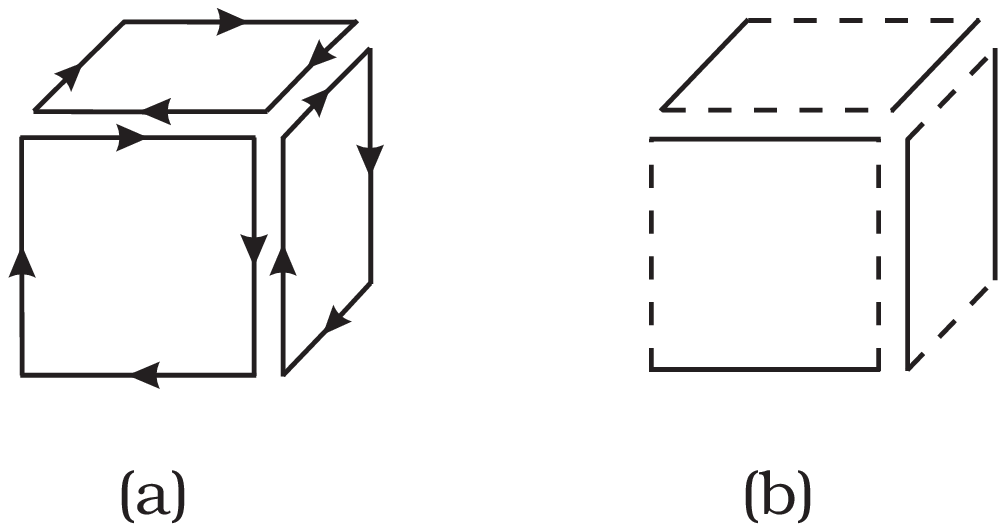,height=6.5cm, width=
,bbllx=38pt,bblly=554pt,bburx=389pt,bbury=756pt, rheight=,rwidth=12cm, clip=}

\vspace*{0.7truein}             
\fcaption{Construction of the monopoles from the field $\theta$ (a), and
of minopoles from the field $\chi$ (b).}
\label{fig:minopoles}
\end{figure}

Since monopoles, which exist in the MaA projection become minopoles in the
MiA projection, then if in the MaA projection the confinement phenomenon is
due to condensation of monopoles (constructed from the field $\theta$), then
in the MiA projection the confinement is due (in some sense) to other
topological objects (minopoles) constructed from the ``matter'' field
$\chi$. It should be stressed that monopoles still exist in
the MiA projection; they can be extracted from the fields $\theta$ in the
usual way, but they are not related to the dynamics. From the point of view
of the initial $SU(2)$ gauge symmetry the fields $\theta$ and $\chi$ are
equal; this fact explains the symmetry between the monopoles and the
minopoles in MaA and MiA projections.

Now we make several simple remarks about various abelian projections from
the point of view of the path integral. We start from the standard partition
function for the $SU(2)$ gluodynamics: $\cZ = \int\, [d U_l]\, e^{-S(U_P)}$.
After the abelian projection we have: $\cZ = \int \, [d\theta] [d \chi] [d
\phi]\, Det^{1/2}(\Delta) \, e^{-S_0 (\theta,\chi,\phi)}$ , where $\Delta$
is the Faddeev--Popov operator. Integrating over the variables $\phi$
we get:

\beq
    \cZ = \int\, [d\theta] [d \chi] \, e^{-S_1(\theta,\chi)}; \label{S1}
\eeq
integration over the variables $\chi$ yields:

\beq
\cZ = \int \, [d \theta] \, e^{-S_2(\theta)}
\eeq
and integration in \eq{S1} over the variables $\theta$ results in:

\beq
\cZ =  \int \, [d \chi] \,  e^{-S_3(\chi)} \label{S3}
\eeq
In the introduction we give several examples which show that in the MaA
the monopoles behave similarly to the Cooper pairs in a
superconductor. This means that:

\begin{itemize}

\item In the MaA projection the action for the monopole {\it
fields} (not the monopole currents) is close to the action of the Higgs
boson in the Abelian Higgs model, the role of the Higgs boson being
played by the monopole field.

\item The quantum theory for this Abelian Higgs model is close to the
(quasi)classical theory.

\end{itemize}

This seems to be related to the fact that $S_2(\theta)$ is
sufficiently local and simple, and, as we have already mentioned, is close
to the action of the compact QED\footnote{$S_2$ can not be {\it equal} to
the QED action, since, {\it e.g.} in QED the asymptotic freedom is absent.}.
Due to the symmetry between MiA and MaA projections the action $S_3$ \eq{S3}
is sufficiently local and simple in the MiA projection\footnote{$S_2$ is a
function of the $U(1)$ invariant loops, constructed from the field $\theta$;
and $S_3$ is the same function of the loops, constructed from the field
$\chi$.}.

\subsection{Hybrids}
MiA and MaA projections are in some sense opposite to each other; there are
infinitely many abelian projections in between the MiA and the MaA
projections.  For these ``intermediate'' projections the action $S_1$
\eq{S1} may be simple. Now the fields $\theta$ and $\chi$ are
important for the dynamics, and the topological objects constructed from
both $\theta$ and $\chi$ may be also important. We call
these objects ``hybrids''. Two examples of hybrids are shown in Fig.2(a,b).
As in Fig.1, the field $\theta$ is denoted by the line with an arrow, and
the field $\chi$ is shown by the solid or dashed line depending on its sign.

{\samepage
The construction of hybrids is similar to the construction of
monopoles and minopoles. Since the angles attached to the plaquettes
are taken (mod $2\pi$), the sum over the phases of the cubes shown in Fig.2
always gives $2\pi\cdot Q$, $Q = 0,\pm 1,\pm 2$, for any $\theta$ and
$\chi$. The charge $Q$, of the hybrid, is invariant under the $U(1)$ gauge
transformations.}

\begin{figure}
\psfig{file=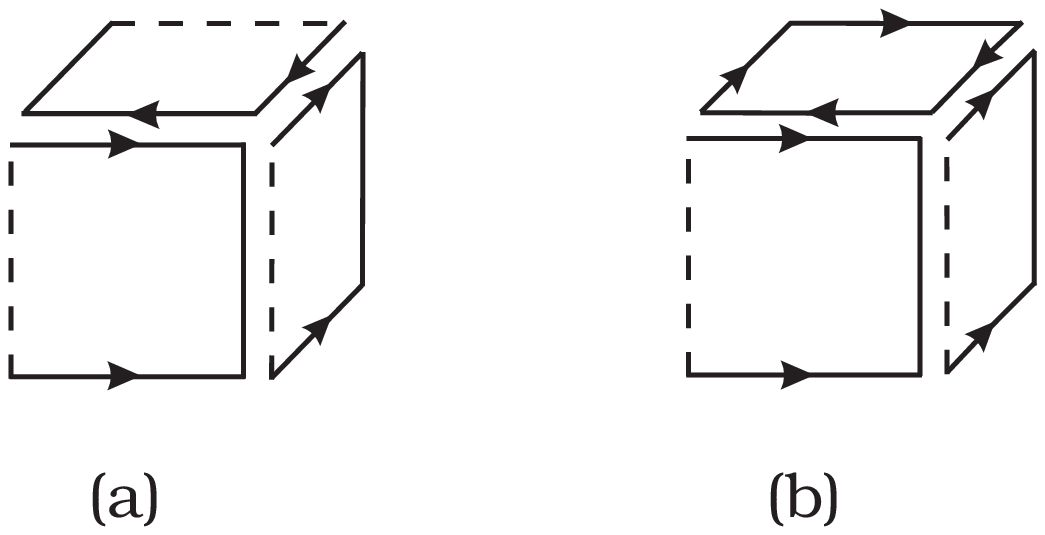,height=6cm, width=
,bbllx=28pt,bblly=544pt,bburx=376pt,bbury=721pt,rheight=1cm, rwidth=,
clip=}
\vspace*{2.8truein}             
\fcaption{Hybrids constructed from the fields $\theta$ and
$\chi$.} \label{fig:hybrids}
\end{figure}

\subsection{Extended Monopoles}

We have widely used the phrases like ``monopoles (minopoles) are
important for the confinement'', but we have not specified the exact meaning
of the word ``important''.  For a quantitative discussion we may use the
following criteria. In the MaA projection the Creutz ratio of the Wilson
loops constructed from the fields $\theta$ gives the string tension which is
close to the full $SU(2)$ string tension $\sigma_{SU(2)}$. In the MiA
projection, in order to get the same value of the string tension, we have
to substitute the loops constructed from the field $\chi$ into the Creutz
ratio. This follows from the exact symmetry $\theta \leftrightarrow \chi$.
Similarly, if we calculate the string tension using monopole currents in
the MaA projection, we have to use the minopole currents in the MiA
projection to get the same result for $\sigma$. Unfortunately, the
confinement scenario is unclear in terms of minopoles; moreover, minopoles
may be, in some sense, lattice artifacts, which do not exist in the
continuum limit.

Still, there exists a possibility that the vacuum is similar to the dual
superconductor in {\it any} abelian projection. The idea is to use the
extended monopoles~\cite{IvPoPo90} defined on the cubes of
size $2^3$, $3^3$, ... . A recent study~\cite{KiMaSu95} of the
energy--entropy balance of the extended monopoles shows that extended
monopoles are important for the dynamics of the temperature phase
transition.

{\samepage
There are many open questions. For example: what is the
action $S^{ext}$ of the extended monopoles; is it simple and/or local?  What
is the dependence of $S^{ext}$ on the type of the abelian projection? If
some extended monopoles are important for the dynamics, what is their size;
is it proportional to the correlation length in the gluodynamics? Is there
any physical meaning in the extended minopoles and the extended hybrids?}

\section{What Kind of String may appear in the $D=4$ Gluodynamics?}

Here we briefly describe the results of our recent
investigation~\cite{AkChPoZu95}. Numerical studies of the lattice
gluodynamics clearly show the formation of a string between the quark and
the antiquark (see, for example, the recent paper~\cite{BaScSc94}). The
string is made of gluons, and is, therefore, the bosonic string. In the
first approximation, the action is proportional to the area of the string
world sheet:

\beq
S = \mu \cdot\mbox{Area} = \mu  \int\, d^2 \sigma \,\sqrt{g}; \label{Sstr}
\eeq
here the standard notations are used: the string world sheet $\tilde
x(\sigma)$ is parametrized by $\sigma_a, \ a = 1, 2$; $g_{ab} = \diff_a
\tilde x_{\mu} \diff_b \tilde x_{\mu}$ and $g = det||g_{ab}||$.

Attempts at numerical simulation of this string
in four dimensions have led to the sophisticated world sheets similar to
``branched polymers''. This is related to the well known
difficulty~\cite{Pol87} in the quantization of the bosonic string in four
dimensions, which can be explained in the following way. For the Nambu-Goto
action \eq{Sstr} we have the Virassoro algebra (algebra of the generators of
the conformal transformations):  $[L_{n}, L_{m}]  =  (n - m) L_{n + m} +
\frac{D -26}{12} (m^3 - m) \delta_{n + m, 0}$; and the last term in the
right-hand side prevents quantization for $D \ne 26$. It
occurs~\cite{PoSt91}, that if we include the additional term in the action:

\beq
S \rightarrow S = \mu \int \, d \sigma \, \sqrt{g}
- \frac{\gamma}{96 \pi}\int \,d \sigma \, (\diff_a \ln{\sqrt{g}})^2,
\label{SPS}
\eeq
the Virassoro algebra takes the form:

\beq
[L_{n}, L_{m}] = (n - m) \, L_{n + m}
+ \frac{D -26 + \gamma}{12} (m^3 - m) \delta_{n + m, 0}. \label{VaSP}
\eeq
If $\gamma=22$, then for $D=4$ the conformal anomaly is absent and the
theory can be quantized. Below we show that this mechanism of
cancellation of the conformal anomaly is natural if one starts from the
field theory.

Consider a $4D$ theory in which strings exist, for example, the Abelian Higgs
theory. Then it is possible to change the field variables to the string
variables, and the Jacobian $J$ appears in the integral:

\beq
\cZ = \int [d \phi] [d A] e^{- S(\phi,A)}
= \int [d\tilde x] e^{- S(\tilde x)}  J(\tilde x)
\eeq

It occurs~\cite{AkChPoZu95} that:

\beq
J(\tilde x)  = \exp\left\{-\frac{11}{48 \pi}\int \,d \sigma \, (\diff_a
\ln{\sqrt{g}})^2 + ...\right\},
\eeq
Comparison with eqs.\eq{SPS}, \eq{VaSP} shows that it is the Jacobian
that gives the term in the string action which cancels the
conformal anomaly in four dimensions! The Jacobian does not depend on the
field theory from which we have started, and therefore the said
mechanism is universal, and can be expected to work in gluodynamics and
chromodynamics.

\section{Acknowledgements}

Our work in 1993 -- 1994 was supported by the JSPS Program on Japan -- FSU
scientists collaboration.  The authors of this talk are very much obliged to
the organizers of this foundation. M.I.P. is much obliged to T.~Suzuki for
many stimulating discussions.

\end{document}